\documentclass{pasj00}

\begin{document}
\SetRunningHead{Oyama et al.}{Proper Motions of the Galactic Center Maser Sources}
\Received{2007/07/09}
\Accepted{2007/08/30 ; version 2.1 Aug. 21, 2007}

\SetVolumeData{2008}{60}{1}
\Published{No.1, Feb. 25, 2008 issue in press}

\title{A Measurement of Proper Motions of SiO Maser Sources 
in the Galactic Center with the VLBA}

\author{Tomoaki \textsc{Oyama},$^{1,2}$ Makoto \textsc{Miyoshi},$^{3}$ Shuji {\sc Deguchi},$^{4,5}$
Hiroshi Imai,$^6$ %
}%
\and
\author{Zhi-Qiang \textsc{Shen}$^{7}$}
\affil{$^{1}$ Institute of Astronomy, School of Science, The University of Tokyo, \\ 2-21-1 Osawa, Mitaka, Tokyo 181-0015}
\affil{$^{2}$ VERA project office, National Astronomical Observatory, \\ 2-21-1 Osawa, Mitaka, Tokyo 181-8588}

\affil{$^{3}$ Division of Radio Astronomy, National Astronomical Observatory, \\ 
2-21-1 Osawa, Mitaka, Tokyo 181-8588}

\affil{$^{4}$ Nobeyama Radio Observatory, National Astronomical Observatory,\\
              Minamimaki, Minamisaku, Nagano 384-1305}
\affil{$^{5}$ Department of Astronomical Science, The Graduate University \\
               for Advanced Studies,\\ Nobeyama Radio Observatory, Minamimaki, Minamisaku, Nagano 384-1305 } 	
\affil{$^{6}$ Department of Physics, Faculty of Science, Kagoshima University, \\
               1-21-35 Korimoto, Kagoshima 890-0065 }

\affil{$^{7}$ Shanghai Astronomical Observatory, Chinese Academy of Sciences \\
     80 Nandan Road, Shanghai, 200030, P.R. China}


\maketitle

\begin{abstract}
We report on the high-precision astrometric observations of maser sources 
around the Galactic Center in the SiO $J=1$--0 $v=1$ and 2 lines 
with the VLBA during 2001 -- 2004.
With phase-referencing interferometry referred to the radio continuum source Sgr A*,  
accurate positions of masers were obtained for three detected objects: 
IRS 10 EE (7 epochs), IRS 15NE (2 epochs), and SiO 6 (only 1 epoch).    
Because circumstellar masers of these objects  
were resolved into several components, proper motions for the maser sources 
were derived with several different methods.
Combining our VLBA results with those of the previous VLA observations,  
we obtained the IRS 10EE proper motion of 76$\pm 3$ km s$^{-1}$ (at 8 kpc)
to the south relative to Sgr A*.  Almost null proper motion of this star 
in the east--west direction results in a net transverse motion 
of the infrared reference frame of about $30\pm 9$ km s$^{-1}$ to the west relative to Sgr A*. 
The proper-motion data also suggests that IRS 10EE is 
an astrometric binary with an unseen massive companion.      

\end{abstract}

\section{Introduction}

It is believed that the center of the Milky Way harbors a super massive black hole \citep{lyn71}. 
Indeed, interferometric observations confirmed the presence of a compact radio source there
\citep{bal74,eke75,lox75}, and the unresolved source was named Sgr A* \citep{bro82}.
In the last decade, dramatic improvements in resolution of infrared instruments enabled to 
detect stellar proper motions under gravitational influence of the central massive object, 
which indicates an enclosed mass of $\approx$ 3.7 $\times$ 10$^{6}$ M$_{\odot}$ \citep{ghe05,eis05}.
An upper limit of 1.8 km~s$^{-1}$ (2$\sigma)$ to the Sgr A* random motion
in the extragalactic reference frame also suggests that it must contain at least 
a mass of 0.4 $\times$ 10$^{6}$ M$_{\odot}$  \citep{rei04}. 
These findings provide compelling evidence that Sgr A* is a source associated with
the super massive black hole.
Because of large extinction at optical wavelengths, observations have been attempted at X-ray, 
infrared, and radio wavelengths \citep{mun05,raf07,she05}. 

Collocation of the positions of the Galactic Center at different wavelengths is
a key issue for the research. The position of
the compact radio source (Sgr A*) appeared to be coincident with an extended 2.2 $\mu$m source 
observed within $\pm$ $1''$ \citep{bec75}. To improve the accuracy of the position of Sgr A*, 
\citet{men97} carried out accurate radio measurements of the positions 
of SiO and H$_2$O masers in red giants 
relative to the continuum emission of Sgr A*; the alignment 
of the radio and $K$-band (2.2 $\mu$m) infrared reference frames achieved an accuracy of $0.03''$.
More precise radio/infrared registration is important for determining relative positions between Sgr A* and
the IR flare \citep{nay05, eck06}, or between Sgr A* and periapses of
stellar orbits \citep{mou05}; in fact,
the periapsis of S0--16 was estimated to be about 45 AU (6 milliarcsec [mas]) away from Sgr A* \citep{ghe05}.
Therefore, attempts were made toward this direction with radio and infrared telescopes \citep{rei03,rei07},
reaching an accuracy of the concordance down to nearly one mas. 

An obstacle for position measurements in radio is a temporal variation of spatial distribution of circumstellar masers.
The brightness distribution of maser emission is not uniform in the circumstellar shell, but rather it is random and patchy. 
SiO masers in a typical mira are emitted at a few stellar radii, i.e., at about $2\times 10^{14}$ cm \citep{cot04}. 
This corresponds to 1.6 mas at the distance of the Galactic Center (8 kpc). 
Therefore a position of the central star estimated from a single maser feature 
may have considerable uncertainty.
Due to its high sensitivity,  past measurements of proper motions of SiO masers
around Sgr A* were made mainly with the VLA (Very Large Array); 
a synthesized beam of 80 mas $\times$ 40 mas towards Sgr A* at 43 GHz, 
though they advocated "combined" VLA and VLBA (Very Long Baseline Array) results 
for only a single-epoch VLBA observation \citep{rei07}.
Though the VLBA is a very powerful tool to investigate a spatial distribution of maser features
(a synthesized beam of 1.8 mas $\times$ 0.5 mas at 43 GHz),
it is limited by the sensitivity \citep{sjo98}. In the case of the SiO masers in the Galactic center sources,
peak line intensities of the SiO $J=1$--0 $v=1$ and 2 transitions are about 0.3 Jy or less.
This is a severe restriction for the VLBI observations, 
which require high S/N ratios of the target maser lines 
in normal data calibration procedures. However,
\citet{deg02} found that SiO maser emission from IRS 10EE occasionally flares; the intensity reached 1.5 Jy 
in the $J=1$--0 $v=1$ line in April--May 2000, 
a factor of 5 stronger than the normal intensity (see \cite{izu98}).
Because infrared monitoring observations demonstrated a periodic light variation of this object \citep{woo98}, 
it is reasonably expected that the SiO maser brightening occurs in every few year.
Therefore, the positional accuracy of IRS 10 EE could greatly be improved if   
the VLBI observations are made at the time of SiO flares.

In this work, we demonstrate abilities and limitations of the VLBA to measure proper motions 
of weak SiO maser sources near Sgr A*. We reports on our few-years attempt 
to improve the positional accuracies 
of IRS 10 EE and two other SiO maser sources in the Galactic center. 
The relative positions of IRS 10EE SiO masers  
with respect to Sgr A* were obtained with unprecedented accuracy. 
However, because maser features have been spatially resolved with the VLBA, 
refined analyses are necessary to extract the position of the central star from the data. Combining them 
with the infrared proper motions of this star, 
we derive the relative velocity of Sgr A* to the infrared reference frame,
which gives a constraint on dynamical motions of various objects surrounding the Galactic Center.

\begin{table*}
\begin{center}
\caption{Summary of VLBA observations.}\label{tab:summary}
\vspace{12pt}
\begin{tabular}{cccccccl}
\hline\hline\\ [-6pt]
Session & nick-  & Date & $v$  & BW &  $	\Delta$$ \nu$  & Acc. period & detected object  \\
name & name  & (yy/mm/dd)  &         &  (MHz)    &     (kHz) & (s)  &  \\

\hline\\
A & 146a & 2001/05/16  & 1,2 & 4$\times$8 & 62.5 & 1.05 & IRS 10EE \\  
B & 146b & 2001/07/31  & 1,2 & 4$\times$8 & 62.5 & 1.05 & IRS 10EE \\ 
C & 146c & 2001/10/05  & 1,2 & 4$\times$8 & 62.5 & 1.05 & failed \\ 
D & 177a & 2003/01/11  & 1,2 & 8$\times$4 & 125 & 0.13 & IRS 10EE \\  
E & 177b & 2003/03/12  & 1,2 & 8$\times$4 & 125 & 0.13 & IRS 10EE, IRS 15NE, SiO 6 \\ 
F & 177c & 2003/10/06  & 1,2 & 8$\times$4 & 125 & 0.13 & IRS 10EE \\ 
G & 131b & 2004/03/08  & 1 & 16$\times$4 & 62.5 & 0.39  & IRS 10EE, IRS 15NE \\  
H & 131c & 2004/03/20  & 1 & 16$\times$4 & 62.5 & 0.39  & IRS 10EE, IRS 15NE \\ 
\hline
\end{tabular}
\end{center}

\vspace{6pt}\par\noindent
Note.---$v$ is the vibrational state of the SiO $J=1$--0 transition involved in the observation. 
Acc. period means an accumulation period.

\label{summary}
\end{table*}
\begin{figure}
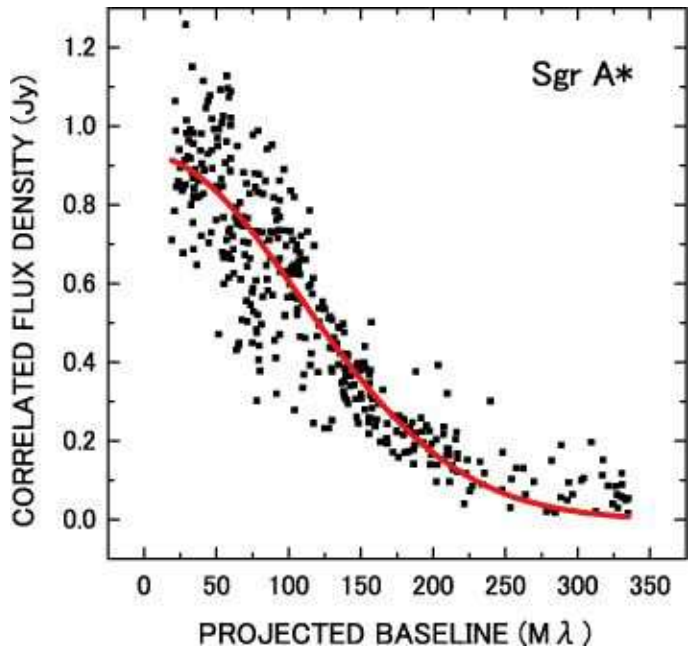

  \begin{center}
    \FigureFile(9cm,9cm){fig1.eps}
  \end{center}
  \caption{Visibility amplitude plotted versus $u$--$v$ distance. The solid curve shows the best 
fitted visibility curves for a circular Gaussian with a zero baseline flux of 0.927 Jy and 
FWHM of 0.712 mas, which is consistent with the size measured at 43 GHz by \citet{lox98}.}\label{sgrastar}
\end{figure}
\section{Observations and data reduction}
The VLBI observations of SiO masers around Sgr A* were conducted between  2001 May and 2004 March
(8 epochs in total), 
using the Very Long Baseline Array (VLBA), a facility of the National Radio Astronomy Observatory (NRAO). 
The present dataset consists of three series of observations, which are summarized in Table \ref{tab:summary}. 
Our primary target stars were chosen from table 2 of \citet{rei03}. They are located within
50$''$ from Sgr A* (R.A. = 17:45:40.0392, Decl. = $-$29:00:28.151 in J2000). 
In this paper, we followed the naming of SiO sources adopted by \citet{rei03}.

The first series of observations was made in 2001. 
The system temperatures (including atmospheric noise) were 100--300 K. 
Received signals were sampled at a rate of 8 M sample 
per second per base band channel (BBC) with 2-bit quantization in only left-hand circular polarization, 
and recorded at the rate of 128 Mbps (16M bit s$^{-1}$ $\times$ 8 BBCs) using the VLBA terminal. In data 
correlation, five different processings were performed, in which the tracking centers 
were set at Sgr A*, IRS 10EE, IRS 7, SiO 6, and SiO 11. 
To observe both $v=1$ and $v=2$ lines, we divided eight BBCs into two groups; 
one group at 42.820587 GHz and another at 43.122027 GHz to cover 
the $v=2$, $J=1$--0 and $v=1$, $J=1$--0 transitions of SiO. 
Each BBC band was divided into 64 spectral channels. Each band yielded 
a spectral resolution of 62.5 kHz, equivalent to 0.435 km~s$^{-1}$
 in velocity, and covered a velocity range of 27.84 km~s$^{-1}$. The four 4 MHz bands of 
one group centered on velocities of $-120$, $-60$, $-20$, and 60 km~s$^{-1}$ 
with respect to the local standard of rest (LSR). In this first series, 
a correlator accumulation time was set to 1.05 s to cover a field of view of 6.2$''$
from the tracking centers. 
Therefore, not only the star located at the tracking center 
but also any other stars were observable within a field of view. For example, IRS 15 NE was 
within the field of view in which IRS 7 was used as a tracking center.
The separation between IRS 15NE 
and IRS 7 (nearest to IRS 15NE among five stars used as tracking center) is 5.9$''$.  
Also, the separations between IRS 28 and IRS 10EE, and between IRS 12N and Sgr A* are 10.4$''$ 
and 7.6$''$, respectively. Therefore, although IRS 28 and IRS 12N were with in the same band, 
significant losses of sensitivity were expected. 

The second series of observation was made in 2003.
The weather conditions were good on 2003 Jan 11 and Mar 3, but
the PT, KP, LA, OV and FD stations were in bad weather on October 6.
To expand the field of view, the correlator accumulation time was set to be 
 0.13 s, yielding the field of view became about 50$''$ from Sgr A*. 
Though the main purpose of this observation was to detect stellar masers as many as possible, 
the number of available BBCs were restricted. 
We used four BBCs of 8MHz each into two groups; one for $v=1$ and another for $v=2$. 
The spectral band was divided into 64 channels, 
yielding a spectral resolution of 125 kHz, equivalent to 0.86 km~s$^{-1}$ in velocity, 
and covered a velocity range of 55.04 km~s$^{-1}$. 
The two BBCs in one group were centered at LSR velocities of $-60$ and $-$10 km~s$^{-1}$. 
As a result, we were able to observe IRS 10EE, IRS 15NE, SiO 6, SiO 11, IRS 17, and SiO 12.

The third series of observations was made on 2004 Mar 8 and March 20 to investigate the structural variation of Sgr A*.
After the observations, three different correlation processings were performed; the tracking centers were set 
at Sgr A*, IRS 10EE, and IRS 9. Two BBCs of a bandwidth of 16 MHz each were used, both of which were set to the $v=1$ line. 
Each BBC was divided into 256 spectral channels, 
giving a spectral resolution of 62.5 kHz, equivalent to 0.43 km~s$^{-1}$ in velocity, 
and it covered a velocity range of 110.08 km~s$^{-1}$. 
The two BBC bands were centered at LSR velocities of $-$6 and 339 km~s$^{-1}$. 
In this series,  the correlator accumulation time was set to 0.39 s, yielding a field of view of 16.7$''$. 
Therefore, not only stars at the tracking center but also IRS 15NE and IRS 28 were 
included within several fields of view. 
The separations between IRS 15NE and IRS 10EE and between IRS 28 
and Sgr A* were 8.8$''$ and 10.9 $''$, respectively. 
As a result, we were able to observe IRS 10EE, IRS 9, IRS 15NE and IRS 28. 
Table 1 gives a detail of the setup for these observations.

\tabcolsep 2pt
\begin{longtable}{lllcccccccccc}
\caption{Position offsets of detected SiO masers from Sgr A*.}\label{tab:position}
\hline\hline
Session & $v$ & No & $\Delta\alpha$ & $\sigma _{\alpha}$ & $\Delta\delta$ &  $\sigma _{\delta}$ & Flux & rms & $N_c$ & $V_{\rm lsr}$ & $ d_{\alpha}$ & $d _{\delta}$ \\
      &   &    & (mas) & (mas) & (mas) & (mas) & {\tiny (mJy km s$^{-1}$)}& {\tiny (mJy km s$^{-1}$)} & & ( km s$^{-1}$) &(mas)&(mas)\\
\hline
\endfirsthead
\hline
Session & $v$ & No & $\Delta\alpha$ & $\sigma _{\alpha}$ & $\Delta\delta$ &  $\sigma _{\delta}$ & Flux & rms & $N_c$ & $V_{\rm lsr}$ & $ d_{\alpha}$ & $d _{\delta}$ \\
      &   &    & (mas) & (mas) & (mas) & (mas) & {\tiny (mJy km s$^{-1}$)}& {\tiny (mJy km s$^{-1}$)} & & ( km s$^{-1}$) &(mas)&(mas)\\
\hline
\endhead
\hline
\endfoot
\hline
\multicolumn{13}{l}{Note--- $v$ is the vibrational state of SiO $J=1$--0 involved in the observation.}\\
\multicolumn{13}{l}{$N_c$ is the number of components detected in channel maps. } \\
\multicolumn{13}{l}{$V_{lsr}$ is the velocity of the strongest component.}\\
\multicolumn{13}{l}{$d_{\alpha}$ and $d _{\delta}$ are standard deviations of the R.A. and Decl. offsets, respectively,} \\
\multicolumn{13}{l}{of the components in channel maps. } \\
\endlastfoot
       \multicolumn{13}{c}{IRS 10EE} \\
A & 1 & 1 & 7684.43 & 0.08 &  4209.21 & 0.25 & 252 & 47 &  2  &  $-$27.9 &   0.24 & 0.20  \\
A & 2 & 1 & 7684.40 & 0.04 &  4209.33 & 0.10 & 556 & 63 &  4  &  $-$28.3 &   0.08 & 0.24  \\ 
B & 1 & 1 & 7684.55 & 0.05 &  4208.85 & 0.17 & 199 & 39 &  4  &  $-$28.7 &   0.05 & 0.09  \\
B & 2 & 1 & 7684.32 & 0.04 &  4208.68 & 0.14 & 108 & 16 &  9  &  $-$28.8 &   0.48 & 0.52  \\
D & 1 & 1 & 7683.55 & 0.06 &  4204.30 & 0.12 & 373 & 73 &  4  &  $-$27.9 &   0.30 & 0.40  \\
D & 1 & 2 & 7681.22 & 0.06 &  4208.67 & 0.19 & 470 & 80 &  4  &  $-$27.0 &   1.31 & 0.56  \\
D & 1 & 3 & 7680.88 & 0.09 &  4209.07 & 0.17 & 673 &103 &  3  &  $-$27.9 &   0.22 & 0.35  \\
D & 2 & 1 & 7683.24 & 0.09 &  4204.18 & 0.14 & 773 &107 &  7  &  $-$27.9 &   0.24 & 0.27  \\
D & 2 & 2 & 7681.20 & 0.09 &  4209.22 & 0.24 & 511 & 99 &  3  &  $-$27.9 &   0.16 & 0.07  \\
D & 2 & 3 & 7680.70 & 0.10 &  4209.96 & 0.22 & 547 &103 &  1  &  $-$27.0 &   0.20 & 0.02  \\
E & 1 & 1 & 7682.08 & 0.02 &  4205.77 & 0.07 & 416 & 41 &  4  &  $-$27.9 &   0.05 & 0.09  \\
E & 2 & 1 & 7683.66 & 0.05 &  4205.00 & 0.14 & 314 & 52 &  5  &  $-$28.8 &   0.18 & 0.22  \\
E & 2 & 2 & 7682.20 & 0.03 &  4205.71 & 0.09 & 525 & 53 &  4  &  $-$27.9 &   0.05 & 0.08  \\
F & 2 & 1 & 7683.81 & 0.04 &  4205.68 & 0.13 & 261 & 39 &  4  &  $-$27.0 &   0.34 & 1.31  \\
G & 1 & 1 & 7684.26 & 0.02 &  4203.17 & 0.05 &1231 & 87 & 13  &  $-$28.7 &   0.12 & 0.12  \\
G & 1 & 2 & 7682.82 & 0.04 &  4203.84 & 0.12 & 570 & 84 &  7  &  $-$27.9 &   0.09 & 0.29  \\
H & 1 & 1 & 7684.34 & 0.03 &  4203.29 & 0.06 & 138 & 16 &  6  &  $-$28.7 &   0.06 & 0.06  \\
           \multicolumn{13}{c}{IRS 15NE} \\
E & 1 & 1 & 1217.24 & 0.03 & 11291.46 & 0.09 & 506 & 60 &  8  &  $-$15.5 &   0.70 & 0.32  \\    
E & 2 & 1 & 1217.36 & 0.05 & 11291.38 & 0.09 & 774 & 83 &  9  &  $-$17.3 &   0.35 & 0.57  \\
G & 1 & 1 & 1216.35 & 0.06 & 11286.67 & 0.15 & 565 & 96 &  5  &  $-$14.2 &   0.10 & 0.23  \\
H & 1 & 1 & 1216.37 & 0.06 & 11286.19 & 0.15 & 370 & 69 &  3  &  $-$14.2 &   0.04 & 0.54  \\
           \multicolumn{13}{c}{SiO 6} \\
E & 1 & 1 & 35234.92& 0.03 & 30662.85 & 0.09 & 542 & 62 &  5  &     52.0 &   0.04 & 0.09  \\
E & 2 & 1 & 35234.83& 0.06 & 30662.37 & 0.15 & 389 & 69 &  4  &     52.0 &   0.92 & 0.37  \\
\end{longtable}
 
 At all of the epochs, the continuum source NRAO 530 was observed approximately every one hour with 
 an integration time of seven minutes to check the system performance and to calibrate the clock parameters.

The data reduction was made in the standard manner with the Astronomical Image Processing System (AIPS).
Initial delay and phase calibrations were made using the calibrator NRAO 530.
Single-band delays were determined with fringe-fitting and then applied to the Sgr A* visibilities.
Accurate delays and rates were determined with global fringe-fitting directly to the Sgr A* visibilities. 
Sgr A* was detected on the baselines at $u$--$v$ distances smaller than about 200 M$\lambda$. 
Figure \ref{sgrastar} shows the visibility amplitude versus $u$--$v$ distance 
for Sgr A* observed on 2003 March 12.
The data was averaged in each single frequency 
band (one channel) after residual delay, rate, phase and amplitude corrections. 
Bad data due to spurious amplitude and unstable phase were flagged out using the AIPS task UVFLG. 
Then, the hybrid imaging process, which involved iterative imaging 
with CLEAN and self-calibration, was made. 
We obtained antenna-based solutions for correcting the atmospheric phase fluctuation. 
These calibration solutions were applied to the all maser objects.
This method is known 
as the "phase-referencing technique", and makes it possible to determine the relative positions 
of maser features with respect to Sgr A*. 
Differential aberrations were compulsorily corrected in the procedure UVFIX. 
After the phase referencing calibration, 
Doppler correction of target maser data was applied using the AIPS task CVEL. Then
we created image cubes of the $v=1$ and $v=2$ masers using the AIPS task IMAGR.
The initial images contained 512 $\times$ 512 pixels with a pixel size of 0.2 mas $\times$ 0.2 mas,
and a taper was applied to the data, which reduced the spatial resolution and made 
the initial search for emission more efficient. The final image contained 
512 $\times$ 512 pixels with a pixel size of 0.05 mas $\times$ 0.05 mas. 

The spatial and velocity structures of maser emission were more or less varying from epoch to epoch. 
The peak positions and flux densities of all maser components were derived 
from the elliptical Gaussian fitting to the components using the AIPS task JMFIT.
If more than one components exist at same velocity channel, 
we estimated the position of each component by a series of single-Gaussian component fits. 
Also, if the structure is too complicated in a channel map, 
we consider an area of a peak flux as a single Gaussian component, 
and estimated the position. 

\section{Results and discussions}

\subsection{Distributions of maser features around stars in the Galactic Center}

\begin{figure*}
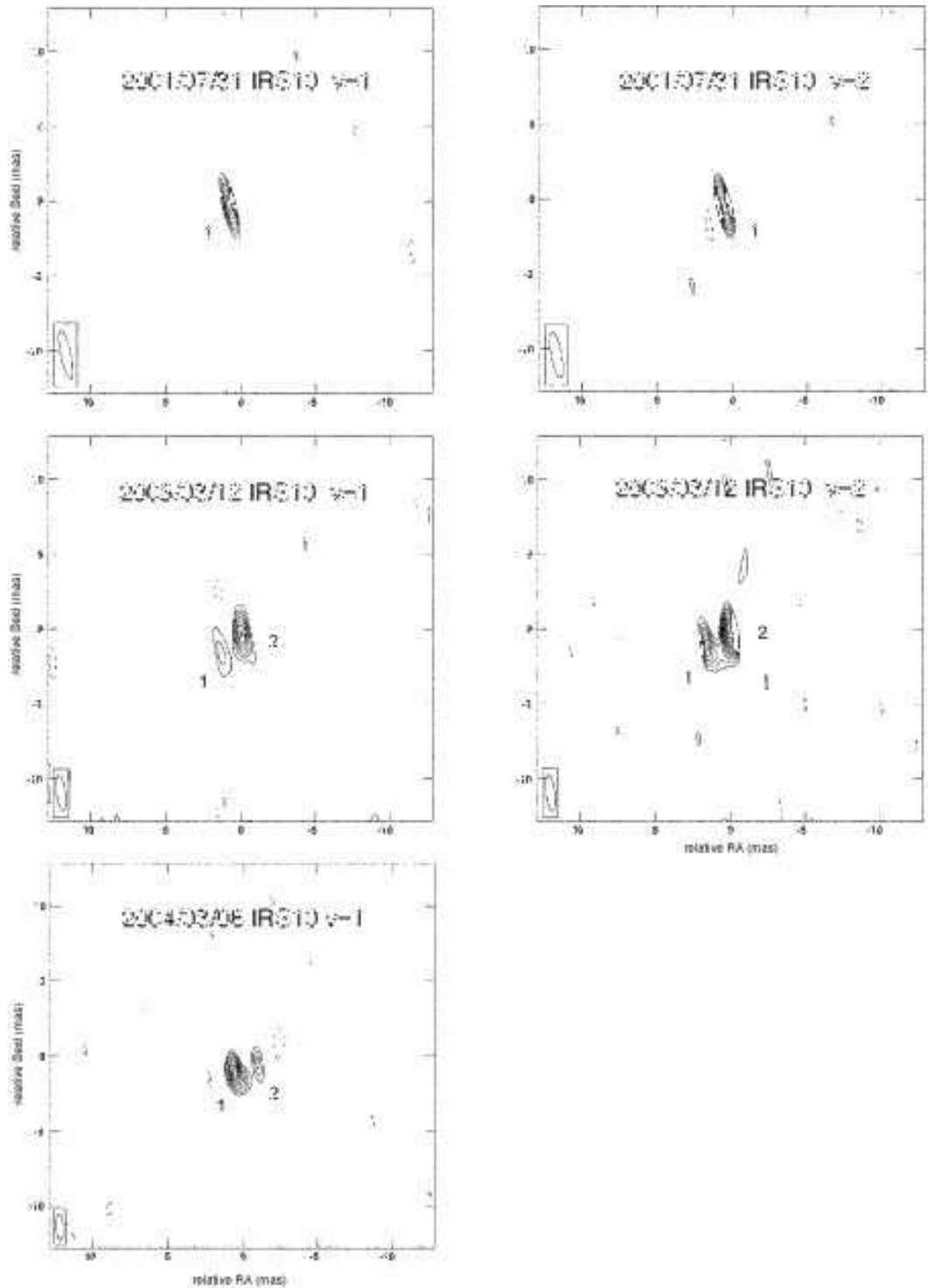

  \begin{center}
    \FigureFile(14cm,19cm){fig2.eps}
  \end{center}
  \caption{Integrated intensity maps of the SiO $J=1$--0 $v=1$ and 2 lines for IRS 10EE
  in July 2001 (top), March 2003 (middle), and March 2004 (bottom). The $v=2$ line
was not observed at the last epoch. The half-power beam is shown at the lower left of each panel. 
Contours are drawn at every noise-level step above 3 times the noise levels,
 20.7  (top left), 25.4 (top right),
18.6  (middle left), 21.9 (middle right), and 26.3 mJy km s$^{-1}$ beam$^{-1}$ (bottom left),
except for the middle-left and bottom panels, which use a step of twice the noise-level. 
 }\label{IRS10EESiOv12}
\end{figure*}

\begin{figure*}
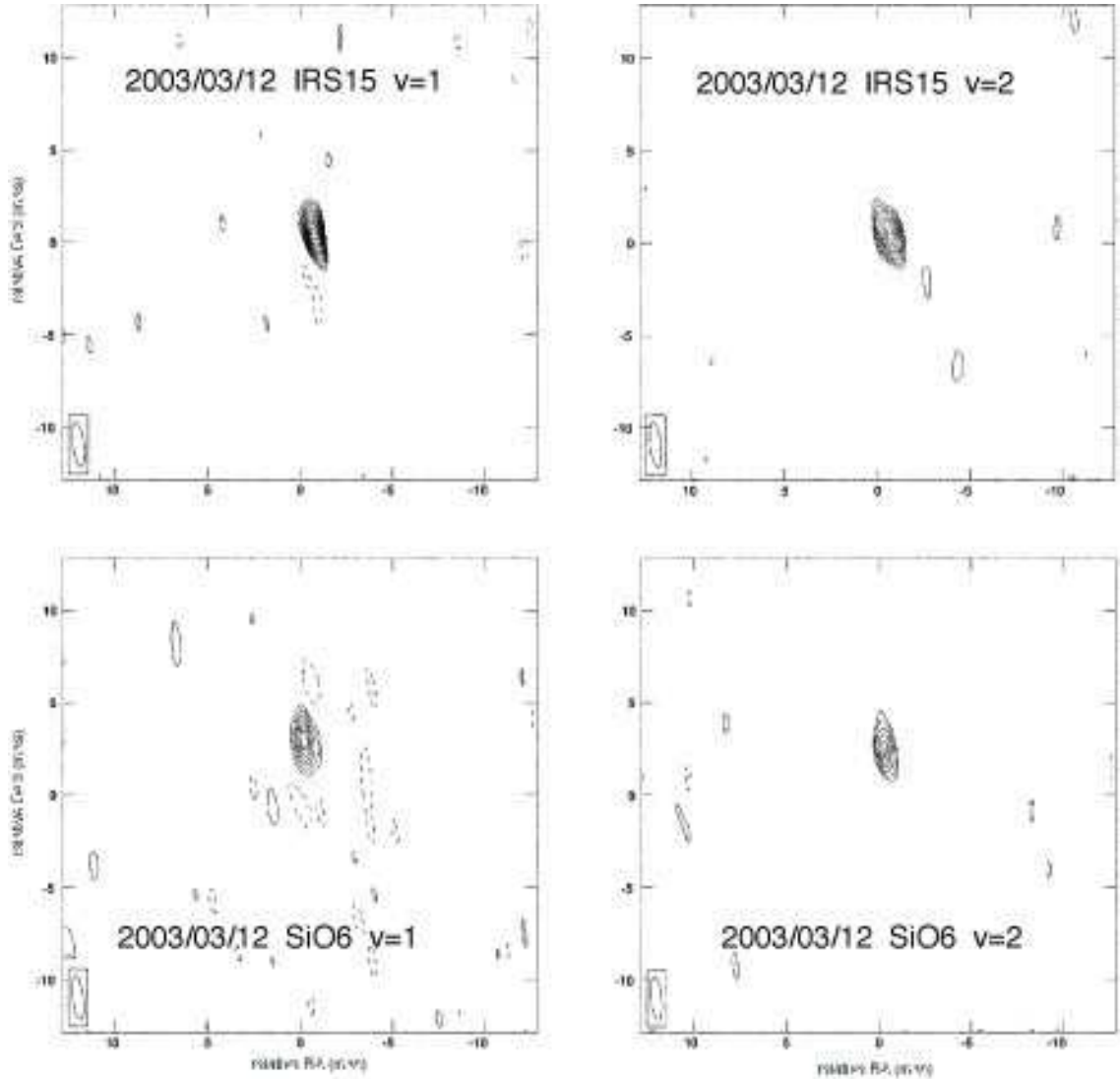

  \begin{center}
    \FigureFile(15cm,15cm){fig3.eps}
  \end{center}
  \caption{Same as figure 2, but for IRS 15 (upper panels)  and SiO 6 (lower panels) in Jan 2003. 
  Contour levels are drawn at every noise-level step above 3 times the noise levels,
  24.7 (top left), 31.0 (top right), 22.3 (bottom left), and 23.2 mJy km s$^{-1}$ beam$^{-1}$ (bottom right), 
  except for the bottom left panel which uses a step of twice the noise levels.
}\label{irs15sio6}
\end{figure*}

SiO masers in the Galactic Center have low flux densities, typically 60--500 mJy. 
Moreover, they are spatially resolved out in the long baselines of the VLBA. 
Therefore, it results in a low detection rate in the present observations. 
Only three sources, IRS 10EE, IRS 15NE, and SiO 6 (a new detection with the VLBA),
 were detected with the shortest 10 baselines at the signal-to-noise ratio (S/N) above 4.
Figure \ref{IRS10EESiOv12} shows integrated intensity maps of IRS 10EE at three epochs
and Figure \ref{irs15sio6} shows those of IRS 15NE and SiO 6 in the SiO $J=1$--0 $v=1$ and 2 lines. 
The integrated-intensity map covers a velocity range of $\sim 8~\rm km~s^{-1}$
 centered at $V_{\rm lsr} \sim$$-$27.0 km~s$^{-1}$ for IRS 10EE, a range of
$\sim 7~\rm km ~s^{-1}$ centered at $V_{\rm lsr}\sim -13.3~\rm km~s^{-1}$ for IRS 15 NE, 
and a range of $\sim 8~\rm km~s^{-1}$ centered at $V_{\rm lsr}\sim$ 52.0 km~s$^{-1}$ for SiO 6. 
These VLBA maps show several maser features distributed in areas with a diameter of 3 mas with 
velocities ranging a few km~s$^{-1}$. 

In IRS 10EE (=OH359.946|0.048), the SiO maser emission was extended 
in an area of 2 mas $\times 4$ mas  
(Figure \ref{IRS10EESiOv12}), though apparent elongation to the north--south is an artifact by the elongated beam. 
It was resolved into several features. 
We simultaneously observed two transitions (SiO $J=1$--0 $v=1$ and 2) except the last two epochs.
The location of the $v=1$ emission well coincided with that of $v=2$.
Furthermore, strong time variation was found. 
The SiO masers flared  on 2003 January (session D).  
The total intensity reached about 1.6 Jy km s$^{-1}$, 
a factor of few stronger than the normal intensity level. 
The masers were split into several features in an extended area 
of about 5 mas $\times$ 3 mas; the north--west features 
were not present at the other epochs (see Table \ref{tab:position}). Because of the large spread of emission at this epoch,
we discarded the data of this session from the proper motion fitting described in the next section.

A scale of 1 mas corresponds to $1.2\times 10^{14}$ cm  at a distance of 8 kpc.  For comparison,
a typical mira variable, TX Cam, exhibits a ring structure of SiO maser emission with a  diameter of 
 $\sim 2\times 10^{14}$ cm at a distance of 450 pc (28 mas; \cite{dia94}), which is comparable with
 the extent of the SiO masers in IRS 10EE. 
In the case of a supergiant, for example, for VY CMa, 
the distribution of SiO maser features is extended in an area of 80 mas ($\sim$ 120 AU) 
at the distance of 1.5 kpc from the Sun \citep{miy03,shi04}, so that it  
would be extended in an area of about 10 mas if placed at the Galactic center. 
The spectral type of IRS 10EE was classified to M0III \citep{fig03} and 
photometric observations suggested that IRS 10EE is a long period variable \citep{tam96,woo98,blu03,pee07}.
The small diameter of SiO maser emission of this star is consistent with these previous findings. 

For the case of the IRS 15NE masers, the SiO emission was slightly resolved 
and extended in an area with a diameter of about 2 mas (upper two panels of Figure \ref{irs15sio6}).
However,  only a single maser feature appeared both at two epochs.
This object was also identified to the AGB star \citep{gen96,gen00},
which is consistent with a measured small maser emission size. 
However, the IR spectrum of this star is blended by radiation from a massive hot star 
(\cite{naj97}, or see a note in table 2 of \cite{blu03}), 
and has been classified to WN8 \citep{pau06,mar07}. 
Therefore, an attention must be paid for the astrometric use of this star.

The SiO 6 masers (lower two panels of Figure \ref{irs15sio6}) were also partially resolved 
over the area of about 2 mas. The extent of the SiO masers suggests that the object is an AGB star. 
OH and H$_2$O masers have been found in this star \citep{lev95,yus95,sjo96,lin92a}. 
Though \citet{lev95} and \citet{yus95} considered that the H$_2$O emission was
 associated with a supergiant or a molecular cloud near Sgr A*,  
subsequent observations \citep{sjo96} revealed that the H$_2$O maser is associated 
with the OH/IR star (OH359.956|0.050)
with a typical doubly peaked spectrum with a velocity separation of 22 km~s$^{-1}$. 
The light variation period of this star is 607 d
([GMC2001] 3--5 or V4928 Sgr; \cite{gla01,pee07}). 
The SiO masers have previously  been detected several times 
at Nobeyama \citep{izu98,miy01,deg02,deg04}, and intensities of SiO masers seem to not vary drastically. 
Our result also confirmed that this is not a supergiant.

\begin{figure}
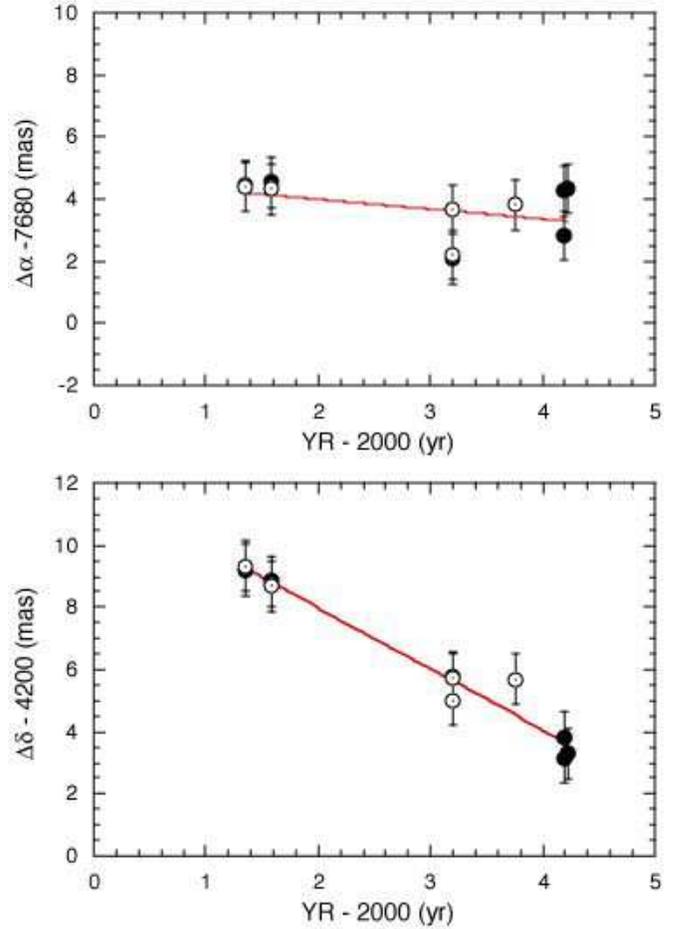

  \begin{center}
    \FigureFile(10cm,10cm){fig4.eps}
  \end{center}
  \caption{Linear fit to IRS 10EE positions relative to Sgr A* in right ascension (upper panel) and in declination (lower panel).
Filled and unfilled circles denote the SiO $J=1$--0 $v=1$ and 2 lines, respectively.
The error bar (1.5 stellar radii; 0.8 mas) is also indicated in each data point.  
The data of session D were discarded. 
  }\label{fitting}
\end{figure}

\subsection{Proper motions of the maser stars}
Stellar masers at the Galactic center are quite faint. 
Therefore, we used the in-beam phase-reference technique together with a long integration time. 
Noise levels in individual channel maps reached $\lesssim 20$ mJy per beam. 
In order to measure accurate proper motions of individual SiO maser stars,
we tried to track the central star position of masers with several different methods.
The first one is to take the average position of all the maser features detected 
with $S/N > 5$ in the integrated intensity map in each epoch. These components are
listed in Table \ref{tab:position}.
Though formal uncertainties of positions for individual components are quite small ($\lesssim$ 0.1 mas),
the data points are scattered in a scale of a few mas. Therefore, it is necessary to re-evaluate
the magnitudes of uncertainties for obtaining the position of the central star.
If the maser features appear randomly in all directions referred to the central star,
the average position must be close to the center position of the star. We estimated
a standard deviation of the position offset of a maser component from the central star to be about 0.8 mas 
(here we assumed the SiO maser feature appears randomly in a flat disk with a radius of 1.6 mas 
(3 stellar radii for a typical AGB star).
Therefore, we added this value to the formal position uncertainty of the individual component.   
Figure \ref{fitting} shows the linear fits to the R.A. and Decl. offsets from Sgr A* as a function of time 
using both $J=1$--0 $v=1$ and 2 lines of IRS 10EE. 
The results obtained with this method are shown in the first row of table \ref{tab:motion}.

The second method is to track a single maser feature.   
For this method to work, we have to select a suitably strong maser feature,
which must last during the observational period. For this purpose,  
we applied the criteria such as
signal-to-noise ratio higher than five in the individual channel map,
detections in the several adjacent channel maps at the same position,
 and detections at three or more epochs at the same velocity. 
With these selection criteria, we determined the center position of the $-27.9$ km s$^{-1}$ component
in sessions B, C, E, and G in both of the $v=1$ and 2 lines for the IRS 10EE masers, 
and computed a proper motion by fitting a straight line to the positions.
The results are listed in the 2nd row of Table \ref{tab:motion}. 

The uncertainties in the second row of Table \ref{tab:motion} are quite small compared with those in the first row
because we used the formal measurement uncertainties of individual-component positions in the second method.
Because a single maser feature is expected to move relative to the central star by an amount 
of about 20 km s$^{-1}$, we need to correct the motion by about 0.52 mas yr$^{-1}$ 
at most for the central star. 
Therefore, taking into account this effect, we can state that
the proper motions obtained by the second method are consistent with the values obtained by the first method,
provided that the $-27.9$ km s$^{-1}$ component is an clump at the south-west edge of the expanding envelope of IRS 10EE,
though this does not clearly reveal from the data.

From OH scattering-size measurements at 1.6 GHz \citep{van92,fra94}, 
angular broadening of a maser feature by interstellar scattering at 43 GHz is expected 
to be up to about 1 mas at the Galactic center.
Therefore, the center of a single Gaussian component is an average position of
several maser features within the same velocity channel, making 
a situation worse, even if we track the same single velocity component. 
\citet{bow01} suggested that the center position of Sgr A* shifts with frequency by 2--5 mas to the east 
between 8.4 and 5 GHz, which was attributed to a difference of source structure between two frequencies
(in Sgr A* and position reference sources). 
However, this can also be attributed to the frequency-dependent refraction of radio waves in plasma
toward the Galactic center. This refraction effect could cause position offsets between radio and optical measurements 
by about 0.1 mas at maximum at 43 GHz, which is not negligibly small. However,
it does not influence the proper motion measurement unless it systematically varies
during the observational time period.

\begin{figure}
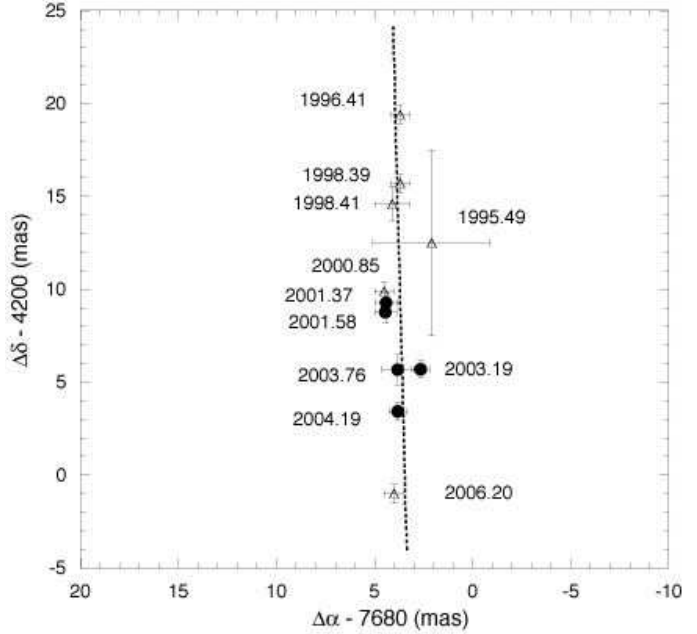

  \begin{center}
    \FigureFile(9cm,9cm){fig5.eps}
  \end{center}
  \caption{IRS 10EE positions relative to Sgr A*. Filled circle and unfilled triangle indicate the average positions of IRS 10EE SiO masers
  obtained at each epoch in this work (with VLBA), and those observed by \citet{rei07} with VLA (except 1996.41 with VLBA), respectively.
  The number indicates the epoch of individual observations. The error bar of the data is also indicated.
  The dotted line shows the best linear fit for combined results.}\label{propmot}
\end{figure}

The third row of Table \ref{tab:motion} gives the IRS 10EE proper motions obtained from Table 2 of \citet{rei07}. 
They reasonably agree with the present results. Note that the uncertainties involved in the first method
are slightly larger than those of \citet{rei07}. This is because we used the position uncertainty 
of 0.8 mas for deducing the central star position. 
Figure \ref{propmot} shows a comparison
of our measured positions (averages) and those of \citet{rei07}. 
Note that the proper motion of IRS 10EE in the east--west direction is negative
in our measurements, while that of \citet{rei07} is slighty positive; 
this can be clearly seen in Figure \ref{propmot}.

\begin{table*}
\begin{center}
\caption{Proper Motions of IRS 10EE, IRS 15NE, and SiO 6.}\label{tab:motion}
 \vspace{12pt}
\begin{tabular}{lcccccccl}
\hline\hline\\ 
Object & $N_e$ & $v$ & $V_{\rm lsr}$ & $\mu _x$ & $\sigma_x$ & $\mu _y$ & $\sigma _y$ &note \\
       &  &   & (km s$^{-1}$ ) &(mas yr$^{-1}$ ) &(mas yr$^{-1}$) &(mas yr$^{-1}$) &(mas yr$^{-1}$)& \\
\hline
IRS 10EE & 6 &1,2 &  all  & $-0.32$ &  0.21 & $-1.96$ &  0.21 & this work ($S/N>5$)  \\ 
         & 4 &1,2 &$-27.9$& $-0.43$ & 0.02 & $-2.25$ &  0.07 & this work (a feature) \\  
        & 6 & 1 & $-27$ & 0.04 & 0.07 & $-2.09$ & 0.07 & \citet{rei07} \\	 
       & 12 & 1,2& all & $-0.06$ & 0.09 & $-2.01$ & 0.09 & combined \\	
IRS 15NE & 3 & 1,2 & all & $-0.92$ &  0.79 & $-4.90$ &  0.79 & this work ($S/N>5$) \\
         & 3 & 1 & $-13.7$ & $-1.60$ &  0.05 & $-5.70$ &  0.06 &  this work (a feature) \\  
         & 6 & 1 & $-12$ & $-1.96$ & 0.07 & $-5.68$ & 0.12 & \citet{rei07}  \\
         & 8 & 1,2 & all & $-1.87$ & 0.10 & $-5.76$ & 0.11 & combined \\
SiO 6  & 4 & 1 &  52 & 2.58 &  0.43 &   1.99 &  0.52 & \citet{rei07} \\	 
       & 5 & 1,2 & all & 2.58 &  0.26 &   1.70 &  0.44 & combined \\
\hline
\end{tabular}
\end{center}
\vspace{6pt}\par\noindent
Note.---$N_e$ is the number of epochs involved.  

\end{table*}

If we combine both results, we expect to obtain better values of proper motions for these stars.
For our data to be compatible with those of \citet{rei07}, we took all the maser components 
with high $S/N$ ($>5$) listed in Table \ref{tab:position} and adopt an uncertainty of 0.8 mas, 
which was used in the first method. 
The combined results are shown at the 4th, 8th and 10th rows in Table \ref{tab:motion}.
Note that the combined results of our VLBA and Reid et al.'s data (mostly with the VLA), do not necessarily give
the middle value, but somewhat different one, because the proper motion is a time derivative between positions. 
The position uncertainties, which are used to weight on the data points, influence the derived values;
here we also added the same uncertainty 0.8 mas mentioned above to the measurement uncertainties of \citet{rei07}.  
The obtained positions of IRS 10EE relative to Sgr A* in unit of arcsec are written 
\begin{eqnarray}
  \Delta \alpha & =  7.68390 [\pm 0.00029]'' - 0.000057 [\pm 0.000097]'' \nonumber \\
&  \times (Y[year] -2000.0),  \\
\Delta \delta & =  4.21198 [\pm 0.00029]'' - 0.002007 [\pm 0.000093]'' \nonumber \\
& \times (Y[year] -2000.0). 
\end{eqnarray}
The combined result offers currently best values of the proper motions of IRS 10EE and we use these values
in the later discussion. Additionally, we noticed in figure \ref{propmot} 
that there still remains a small periodic shift of the star position from the best fit line;
to the east during 2000--2001, and to the west during 2003, returning back to the east in 2004--2006. 
This shift seems to be a motion due to a binary system. 
We discuss on this issue in Appendix 1. The analysis demands further corrections
in proper motions, but they are negligibly small. Therefore, we discarded them in the later discussions.
 For IRS 15NE, we get
\begin{eqnarray}
  \Delta \alpha & =   1.22364 [\pm 0.00039]'' - 0.001873 [\pm 0.000104]'' \nonumber \\
  & \times (Y[year] -2000.0), \\
  \Delta \delta & =11.31044 [\pm 0.00041]'' - 0.005761 [\pm 0.000111]'' \nonumber \\
  & \times (Y[year] -2000.0),
\end{eqnarray}
We can compare these results with nominal NIR positions of IRS 10EE and IRS 15NE, 
and pin-point the Sgr A* positions on their NIR images. 
However, because the difference between \citet{rei07} and the combined result is 
quite small ($\lesssim$ 0.1 mas yr$^{-1}$) compared with the position uncertainties of stars 
in the infrared (typically more than 10 mas),  
we cannot effectively improve the Sgr A* position in the high resolution NIR images better 
than \citet{rei07} did. 

The proper motions of several hundred stars have been measured on the NIR images.  
They are individual star motions relative to the average motion 
of the $\sim 400$ ``well-behaved'' stars within $\sim 30''$
of the central cusp \citep{rei03,rei07}.  
Referring to the IRS 10EE proper motions, we can derive the motion of Sgr A* 
relative to this NIR reference frame, i.e., relative to the central cusp. From the IR proper motions 
($\mu^{\rm IR}_x$ and $\mu^{\rm IR}_y$ in table 4 of \cite{rei07}), we can compute the relative motion as
($\Delta\mu _x$, $\Delta\mu _y$)= ($0.79 \pm 0.24$,  $0.08\pm 0.27$) [mas yr$^{-1}$] using IRS 10EE,
and  ($-0.53\pm 0.48$, $-0.53\pm 0.37$)  [mas yr$^{-1}$] using IRS 15NE.
The proper motion in R.A. direction obtained from IRS10EE 
is significantly larger than the standard errors (3.3 $\sigma$ significance). 
The same motions obtained using IRS15NE involve larger errors, so that they are discarded.
We conclude that the relative motion of the Sgr A* to the nuclear star cluster is 0.79 mas yr$^{-1}$
to the east (or $29.9 \pm 9.1 $ km s$^{-1}$ at a distance of 8 kpc).
\citet{rei07} also derived the RA relative motion of $24\pm 9$ km s$^{-1}$ from
their data (2.7 $\sigma$), which is consistent with this work. 
It should be noted, however, that the value strongly depends on the accuracy of proper motion measurements of stars
on the NIR images and sampling of "well behaved" stars. It is supposed that blending and binarity 
of reference stars in the dense central star cluster may strongly affect to the results,
especially the motion of IRS 10EE in this work.
Therefore, above conclusion is somewhat provisional, and future NIR high-precision astrometry 
can examine the validity of above result.

\subsection{A speculation on the anomalous motion of Sgr A*} 

The average of the velocities of $\sim 400$ stars in the central cusp was estimated 
in an accuracy of about 5 km s$^{-1}$ 
(a statistical deviation due to random motions of the reference stars; \cite{rei07}), which is much
larger than a brownian motion of the central black hole in the central cusp ($\sim 0.2$ km s $^{-1}$; \cite{mer07}). 
Therefore, above mentioned anomalous motion of Sgr A* with respect to the NIR reference frame
of about 30 km s$^{-1}$ is considerably larger than the statistical deviation, and needs to be
explained.  Here, we give a speculation on this anomalous motion. 

Suppose that a majority of stars in the Galactic center rotate around the barycenter 
of the Galaxy (here, Sgr A*) in the same direction. 
If a sampling is complete, the average transverse velocity
of these stars with respect the barycenter would be zero due to point symmetry. 
However, if thick dust is present near the Galactic center,
magnitude-limited sampling of stars is biased towards the front-side stars of the Galactic center,
because the stars behind are fainter. In fact, \citet{dav98} and \citet{ale99}
studied K-band extinction within a few arcminutes of Sgr A* and found that
the increase of extinction by about $\Delta A_K \sim 0.3$ toward Sgr A* (down to nearly 30$''$ radius).
A simple calculation based on the $r^{-2}$ cusp density law leads about 39\% of stars seen towards Sgr A* 
within a projected distance of 1.5 pc ($\sim 40''$)
to be actually located at radii larger than 1.5 pc from Sgr A*.  
Therefore, the sampling of these distant stars are possibly biased to the near side
due to extinction by thick dust located near Sgr A*.
If we assume a circular rotation of stars around Sgr A* in the same sense,
this bias results in the preferential selection of the stars moving in one direction.
An estimation of the magnitude of this effect gives a net motion of about 
29 km s$^{-1}$ [$\approx 120$ km s$^{-1}$ $\times 0.39\times 0.5/(0.61+0.5\times 0.39$)] ; 
here all stars behind Sgr A* at radii over 1.5 pc are assumed to be not seen 
and the average velocity of stars within 1.5 pc is zero).
Though the value is strongly dependent on the geometry and velocity distribution of the stream,
it yields an acceptable magnitude of net motion of NIR reference frame with respect to the barycenter of the Galaxy. 
The observed direction of the motion of the NIR reference frame is to the west referred to
Sgr A*. Therefore, the direction of the stellar stream has an angle of
about 60 degree from the Galactic plane and the sense of rotation is opposite to the Galactic rotation.

It has been discussed that an intermediate-mass black hole (IMBH; $\sim 10^3$ -- $10^4 M_{\odot}$) 
may strongly influence on the dynamics of the nuclear star cluster containing a super massive black hole 
\citep{por06,ber06,mat07}. Numerical simulations found that the IMBH alters eccentricities 
of stellar orbits and, as a result, alters
the power-law index of the central cusp considerably. 
IMBH may also be responsible for ejecting hypervelocity stars from the central star cluster \citep{bau06}. 
\citet{eis05} and \citet{pau06} found two young-star disks with total masses of $\sim 1.5\times 10^4\ M_{\odot}$
and ages of about 6 Myr; one rotates clockwise, and another counter-clockwise.
Therefore, dynamical structure of stars at the inner few arcsec seems considerably complex.
Because the 400--600 IR position reference stars, which were used for estimating the average motion 
of the nuclear star cluster, are mostly located out of Sgr A* by more than a few arcsec, 
they are not much affected by the IMBH which is supposedly at a few arcsec from the Galactic center
(e.g., IRS 13E; \cite{sch05}). Moreover, the observed power-law index of the cusp density between 4$''$ and 100$''$
is approximately 2 (for example, see figure 7 of \cite{gen03}). This is close to the
steady-state power low index of 7/4 of the cusp star density 
with a super massive black hole  (e.g., \cite{bah76}), 
indicating that the IMBH does not influence to the motion of stars outside a few arcsecond. 
Therefore, it is unlikely that the streaming motion associated with the NIR reference frame is
caused by IMBH. 
The total mass within a radius of 15 pc is about $3\times 10^7\ M_{\odot}$ 
(for example, see Figure 6 of \cite{deg04}), which is by about one order of magnitude larger
than the mass of the central black hole. 
Therefore, the streaming motion of stars discussed here must be a dynamical property related 
to the central core of the inner Galactic bulge.     

\section{Summary}
The 8-epoch VLBA observations of maser sources in the Galactic center were made
in the SiO $J=1$--0 $v=1$ and 2 lines at $\sim$43 GHz during 2001 May -- 2004 March. 
The SiO masers in IRS 10EE had been detected through all of the epochs, 
while those in IRS 15NE and SiO 6 were detected only at several epochs. 
The radio continuum source Sgr A* were also observed 
within the same single-dish telescope beam.  
Accurate astrometric positions of the SiO maser features relative to Sgr A* 
were obtained at each epoch with the simultaneous VLBI imaging of the maser -- continuum source pairs.  
However, because of the spatial distribution of maser features 
in the circumstellar envelope, 
the derivation of the accurate values of proper motions for the maser sources was not straightforward.  
The proper motions were derived with two different methods: one,
using the average positions for all the detected features with high signal-to-noise ratios, 
and the other, using the positions of long-lasting features; both methods gave very similar values. 
Thus, the precise measurements of proper motions of maser sources relative to Sgr A* were made.
Combining our VLBA positional data with the former measurements by VLA, 
and with NIR proper motion of IRS 10EE, 
we found an unexplained net motion of Sgr A* relative to the nuclear star cluster 
of about $30\pm 9$ km s$^{-1}$. 

\

The authors thank Prof. Y. Sofue for comments on this work.
They also thank Prof. H. Kobayashi and N. Kawaguchi for their encouragements though the period 
completing this work.  This is a part of thesis of T. O. for the partial fulfillments of 
Ph. D. degree for the University of Tokyo.
This research was partially supported by Grant-in-Aid for Scientific Research (C15540242) 
of Japan Society for Promotion of Sciences.

\section*{Appendix 1. IRS 10EE binary system}

\begin{figure}
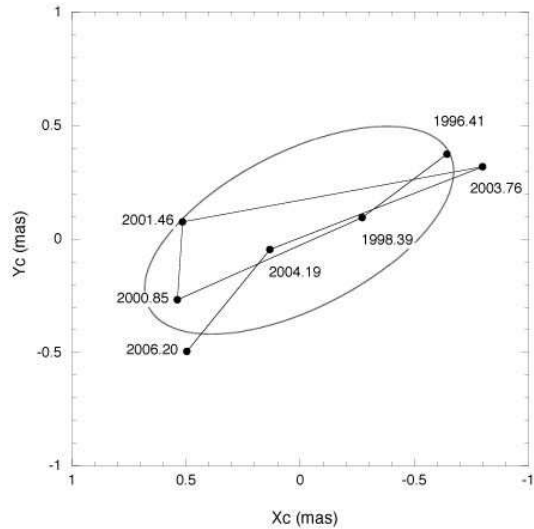

  \begin{center}
    \FigureFile(7cm,7cm){fig6.eps}
  \end{center}
  \caption{Orbit of IRS 10EE.  The filled circles in the line graph 
indicate annually averaged positions of SiO masers after linear motions are subtracted. 
The best-fit ellipse is shown in solid curve.  
}\label{binary}
\end{figure}

We noticed that the positions of IRS10EE in Figure \ref{propmot} deviate systematically 
from the regression line to the east during 2000--2001, to the west during 2003, 
and returning back to the east in 2006. 
It looks like a wavy motion of an astrometric binary star with an unseen companion. Therefore,
it is worth to pursue a possibility of the binary system with a period of about 6 yr. 
In fact, \citet{pee07} found that the X-ray source CXOGC J174540.7$-$290024,
which was listed in table 2 (on-line material) of \citet{mun03}, 
is located within 0.8$''$ of IRS 10EE (their IRS 10*), and concluded that IRS 10EE is a binary system
involving a white dwarf. This source has X-ray luminosity $L_X=1.1 \times 10^{32}$ erg s$^{-1}$,
which is considerably strong for a AGB star with a white dwarf companion
($\sim 10^{29}$ erg s$^{-1}$). 


\begin{table}
\begin{center}
\caption{Orbital elements for the IRS 10EE binary system }\label{tab:binary}
 \vspace{12pt}
\begin{tabular}{lcc}
\hline\hline\\ 
Element & unit & value  \\
         &      &        \\
\hline
$\Delta \alpha _0$ & mas & 0.137 ($\pm 0.186$) \\
$\Delta \delta _0$ & mas & $-0.115$ ($\pm 0.033$) \\
$\Delta \mu _{x} $ & mas yr$^{-1}$ &  $-0.029$ ($\pm 0.065$) \\
$\Delta \mu _{y} $  & mas yr$^{-1}$ &   0.013 ($\pm 0.012$) \\
$T$ & yr &  5.65 ($\pm$ 0.12)  \\
$a_1$    &  mas & 0.73 ($\pm 0.41$)  \\ 
$b_1$    &  mas & 0.35 ($\pm 0.10$) \\ 
$e_1$    &      & 0.87 ($\pm 0.24$)  \\
$i$    &  degree & $\approx 0$ ($\pm 4$) \\
\hline
\end{tabular}
\end{center}

\vspace{6pt}\par\noindent
Note.---$\Delta \alpha _0$, and $\Delta \delta _0$ are the corrections for the center positions of the ellipse, and
$\Delta \mu _{x}$ and $\Delta \mu _{y}$ are corrections for the proper motions, relative to
the quantities given in equations (1) and (2). Quantities, $T$, $a_1$, $b_1$, $e$, and $i$
are the period, semi-major and semi-minor diameters of the primary, ellipticity, 
inclination of the orbit to the sky plane, respectively. 
\end{table}

Therefore, it is interesting to attribute the residuals from the best-fit line [equations (1) and (2)]
 to a binary motion of IRS 10EE, and to check whether or not the derived orbital parameters 
of the binary system are reasonable. 
However, a strong restriction for a binary model is obtained 
from the radial velocity of IRS 10EE; the radial velocity of SiO masers of IRS 10EE ($V_{lsr}\sim -27$ km s$^{-1}$ ) 
did not change more than a few km s$^{-1}$ in the past 10 years. This fact gives a severe condition on
the binary orbit; it must be almost face-on to us.
Note that this binary model is valid only when the apparent motion revealed in figure \ref{propmot}
is attributed to a binary motion. However, we cannot completely deny the other possibility 
that it is due to a systematic periodic variation in SiO maser emission distribution in the
circumstellar envelope of IRS 10EE.

\begin{figure}
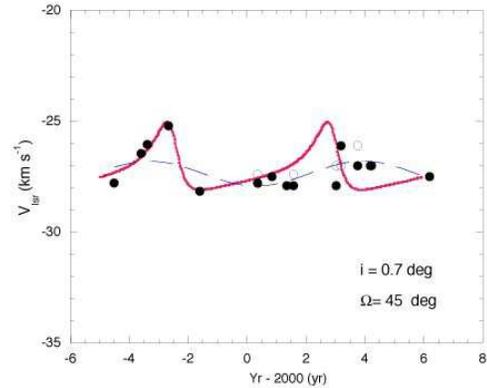

  \begin{center}
    \FigureFile(7cm,7cm){fig7.eps}
  \end{center}
  \caption{Radial velocity variation of IRS 10EE SiO masers. The filled and unfilled circles indicate the
observed peak velocities of the SiO $J=1$--0 $v=1$ and 2 lines, respectively, in the present paper
and in the past literatures. The thick curve indicates the velocity variation expected from the elliptical orbit 
given in Table 4 with an inclination angle of $i=0.7^{\circ}$. 
The broken curve with smaller amplitude is a sine-curve fit for comparison. 
}\label{velocity-variation}
\end{figure}

Though the procedure obtaining orbital elements of a binary system has been well established
(for example, see a brief introduction by \cite{asa07}), it is not straight forward in the present case
because of large positional errors of the central star involved in SiO maser astrometry.
The best-fit line obtained in section 3.2 
does not represent the trajectory of the barycenter of the binary system, 
but merely an approximate average position of the primary (here IRS 10EE) of the binary system. 
It has to be corrected for the elliptical orbit
of the primary. In order to obtain the corrections for the motion of the center 
of the ellipse of the primary, we again fit the R.A. and Dec. residuals of IRS 10EE from the
average motion [equations (1) and (2)] by introducing the linear shift plus a $sin$ curve (as a function of time), 
and determined the period, corrections for the ellipse center, and  corrections for proper motions. 
After correcting these quantities for the center of the ellipse, 
we plotted the residual $X_c$ and $Y_c$ in Figure \ref{binary}, and fit an ellipse to these points 
using the formula of \citet{asa07}. Then, we obtained the orbital elements for the primary,
which are listed in Table \ref{tab:binary}. Because of the above mentioned restriction 
on the orbit being face-on (i.e., $i\approx 0$), the ellipse we obtained on the sky plane must be very close 
to the real binary orbit.
Alternatively, the inclination angle $i$ can be determined by the condition of constant areal velocity 
(the Kepler's second law; \cite{asa07}).
However, such a procedure did not give a reasonable inclination angle in the present case 
because of large scatter of the data points. 
Figure \ref{velocity-variation} shows
the observed radial velocities of SiO emission peak of IRS 10EE in the literatures
(e.g., see \cite{izu98,deg02}) and in the present paper. 
The very small variation of radial velocities ($\Delta V_{\rm lsr}\sim 3$ km s$^{-1}$)
seems to be fit by the elliptic orbit model given in table 4 
with an inclination angle of $i=0.7^{\circ}$, though such a small variation of peak velocities 
can also be attributed to a variation of velocity fields in the expanding envelope without a binary model.  

From the obtained orbital elements, we can compute an approximate mass of the unseen companion using
the Kepler's third law,          
\begin{equation}
m_1 + m_2 = a^3/T^2   
\end{equation}
where $m_1$ and $m_2$ are masses of primary and a companion 
(in unit of solar mass) and $T$ is the orbital period
(in unit of year), and $a$ (in unit of astronomical unit) is the mean separation of the binary, which can be written 
with the semi-major diameters of elliptic orbits of two stars as $a = a_1 + a_2$. Here,
they are constrained by $a_1\ m_1 = a_2\ m_2$.  In order to determine the mass 
of the companion, we have to give a mass of the primary (in addition to $a_1$ and $T$).
The mass of IRS 10EE (primary here) is deduced to be 3 -- 6 $M_{\odot}$, because it is a relatively massive AGB star.
From equation (5), we obtain the mass of the companion to be in a range of 13.1 -- 16.2 $M_{\odot}$.
Because this massive companion is not seen on infrared images, it cannot be a main sequence star; 
it must be a black hole (a white dwarf or neutron star is excluded because of the large mass). 
The mean separation between two stars is in the range of
1.1 -- $1.2\times 10^{14}$ cm, and the mean orbital speed of IRS 10EE is about 30 km s$^{-1}$. 
The periastron distance between two stars is in the range of 1.8 -- $1.9\times 10^{13}$ cm, 
indicating that the secondary grazes the extended atmosphere of the primary ($r_e\sim 1.5 \times 10^{13}$ cm for M0III)
at the periastron passage, when a drastic event may happen. 
However, the periastron distance strongly depends on the uncertainty of $1-e$ of the orbit, 
and it may vary by a factor of 3.  

Estimated values of the parameters for the IRS 10EE binary system seem 
to stay within tolerable ranges, though they are slightly extreme 
in a sample of stellar-mass  black hole binaries \citep{{mcc03}}.  
Therefore, under the condition that the observed nearly null radial velocity variation of SiO masers
indicates the face-on geometry of the orbit, 
we conclude that IRS 10EE is an astrometric binary possibly with a black hole companion.




\end{document}